\documentclass[preprint,12pt]{elsarticle}

\usepackage{amssymb}
\usepackage{amsthm}

\usepackage{graphicx}
\usepackage{subcaption}
\usepackage{caption}

\usepackage{xcolor}
\usepackage{mathtools}

\journal{Engineering Fracture Mechanics}

\begin{document}

\begin{frontmatter}



\title{Finite element simulation of the structural integrity of endothelial cell monolayers: a step for tumor cell extravasation}


\author[label1]{A. Nieto}
\author[label1]{J. Escribano}
\author[label2]{F. Spill}
\author[label1]{J.M. Garcia-Aznar}
\author[label1]{M.J. Gomez-Benito}

\address[label1]{Multiscale in Mechanical and Biological Engineering (M2BE), Department of Mechanical Engineering, Arag\'{o}n Institute of Engineering Research (I3A), University of Zaragoza, Zaragoza, Spain}
\address[label2]{School of Mathematics, University of Birmingham, Birmingham, B15 2TT, UK}

\begin{abstract}

Cell extravasation is a crucial step of the metastatic cascade. In this process, the circulating tumor cells inside the blood vessels adhere to the cell monolayer of the blood vessel wall and passes through it, which allows them  to invade different organs and complete metastasis.  In this process, it is relevant to understand how the adhesions between cells that form the endothelial monolayer are broken, resulting in intra-cellular gaps through which tumor cells are able to extravasate the blood vessel wall.

Within this process, we focus on studying the dynamics of cell-cell junctions rupture produced in the endothelial monolayer by the effect of Calcium waves. The regulation of this monolayer is of vital importance, not only in metastasis, but also in diseases such as pulmonary edema or atherosclerosis.

In order to understand this rupture dynamics in greater depth, we propose a hybrid model that simulates endothelial cells as \textcolor{black}{an} elastic material and cell-cell adhesions of the monolayer by means of a catch bond law.

We study the effects that the cell contraction caused by a Calcium wave presents on the endothelial monolayer depending on the diameter of the blood vessel. For this purpose, we develop a three-dimensional model to study the effect of the different blood vessel diameters.

The results indicate that there are greater tractions on the joints located in vertices common to several cells. This led to the formation of openings in the endothelial monolayer, through which extravasation of tumor cells could occur. For the different geometries studied, no significant effect of the blood vessel diameter on the rupture of the adhesions of monolayer is observed.

\end{abstract}


\end{frontmatter}



\section*{Acronyms}

\begin{tabular}{p{35mm} c p{120mm} }
AJs & --- & Adherens Junctions.\\
VE-Cadherin& --- & Vascular endothelial cadherin.\\
HUVECs & --- & Human umbilical vein endothelial cell.\\
ECM & --- & Extracellular Matrix.\\
FAJs & --- & Focal Adherens Junctions.\\
FAs & --- & Focal Adhesions.\\
LAJs & --- & Linear Adherent Junctions.\\
RSFs & --- & Radial Stress Fibers.\\

\end{tabular}

\section{Introduction}

Cell extravasation of cancer cells is part of the metastatic cascade.  In this process, cancerous cells depart from the primary tumor and are able to intravasate into the blood flow. Once inside, these cells move through the blood until they arrest in the blood vessel wall, which consists of a monolayer of endothelial cells (i.e. the endothelium), and extravasate in order to colonize a new tissue or organ.  This is possible due to the creation of gaps in the cellular monolayer of the blood vessel. The endothelium is formed of a thin lamina of endothelial cells that are separated from the outer ones by an elastic membrane. The endothelial cells are joined by protein complexes such as vascular endothelial cadherin (VE-Cadherin), Nectin, PECAM, etc \cite{Dejana2008}, that form a dynamic structure in which adhesions are broken and rebuilt all the time. This dynamic is responsible for the creation of the openings through which the cancer cells can extravasate \cite{Escribano2019}. Cell transmigration involves the generation of mechanical forces through the actomyosin cytoskeleton and the deformation of the endothelium whose mechanical properties provide passive resistance \cite{CAO20161541}. Different studies have revealed that endothelial \textcolor{black}{monolayer} properties are crucial in gap formation \cite{Oldenburg2014} and that higher levels of junction stiffness can reduce paracellular extravasation \cite{Martinelli2014}. This suggests the importance of analyzing the main drivers behind cell-cell adhesion.

The formation and maintenance of tissues are not only driven by \textcolor{black}{chemical} but also \textcolor{black}{mechanical} processes. The endothelium is subjected to a dynamically changing mechanical environment (i.e. oscillations in blood flow-rate and pressure conditions) which can induce strains in the lining of arteries. Moreover, tangential cyclic forces maintained in time can also change the mechanical properties of the tissue itself \citep{Oldenburg2014}. The endothelium additionally acts as a barrier. It must allow immune system cells to go through while blocking pathogens, blood or tumor cells. One of the key points in this mechanism is the contraction of the actomyosin cytoskeleton of endothelial cells, although there are many points open to be investigated.

Despite the clinical relevance of metastasis and tumoral extravasation, little is known about the mechanical environment that regulates it \cite{Kumar2009a}. The idea that cellular contraction can impact the barrier function of the endothelial monolayer was pointed out in the 70s when several proteins involved in this process were identified \citep{Schnittler2016}. In the last four decades, it has been proven that this mechanical contraction is key in both the mechanical behavior and the development of cells and tissues \cite{Wozniak2009}. In order to improve the understanding of this process, two main modeling approaches are adopted: \emph{in vitro} and \emph{in silico} models. \emph{In vitro} models reproduce the simplified conditions of cell extravasation by controlling the main elements involved (i.e. cells, extracellular matrix (ECM), growth factors). Although the microenvironment cannot be perfectly controlled in these models, they allow comparative analysis and the study of cell-cell and  cell-extracellular matrix interactions \cite{Uygur2011}. Within \textcolor{black}{these} \emph{in vitro} \textcolor{black}{models}, microfluidic devices allow greater control of the microenvironmental variables. They have been used both for the study of cell migration and the intravasation of mechanical barriers \cite{Mak2013a} and cell extravasation \cite{Jeon2013}. Funamoto et al. \cite{Funamoto} studied the behavior of endothelial monolayers under hypoxic conditions using microfluidic channels and Valent et al. \cite{Valent2016} measured contraction forces of human umbilical cord endothelial cells (HUVECs) using traction force microscopy. Few numerical models have focused on the simulation of cell extravasation \emph{\textcolor{black}{in silico}}. Chen et al. \cite{Chen2018} simulated cancer cell deformation during intra- and extravasation, they considered both chemotactic and durotatic factors. Ramis-Conde et al. \cite{Ramis-Conde2009} created a mathematical model focused on intravasation of tumoral cells with a multiscale focus taking into account both intra- and inter-cellular proteins and the shape of the blood vessel. Regarding cellular monolayer, Gonz\'{a}lez-Valverde and Garc\'{i}a-Aznar \cite{Gonzalez-Valverde2017} created a hybrid model focused on the simulation of the dynamic of cellular monolayer combining numerical discrete and continuous models. Recently, Escribano et al. \cite{Escribano2019} created a discrete two dimensional (2D) model in which they analyzed the dynamic behavior of an endothelial monolayer. They studied how different mechanical factors influence cell-cell adhesion rupture and the consequent generation of gaps in the monolayer.

To regulate barrier function, permeability factors influence the following elements of the cellular structure: cell-cell adhesion complexes, cytoskeleton and integrin-ECM adhesions. First, cell-cell adhesion complexes are the last line of defence against vascular permeability. VE-Cadherins are chemically modified through phosphorylation, thus impacting the stability of endothelial junctions \cite{Wong2000}. Phosphorylation regulates the interplay between the VE-Cadherin complex and other complexes, which at last determine the mechanical strength of cell-cell adhesions. Second, the myosin motor activity in the cytoskeleton is key when permeabilization occurs \cite{Brieher2013}, stabilizing cell structure when it is close to actin bundles and networks \cite{Shewan2005}; whereas, when actin is mainly found on radial stress fibers (RSFs), instability and traction forces increase in the cytoskeleton \cite{Ando2013}. Changes in actin bundles seem to influence the local structure and cell-cell adhesion forces. Changes in actin bundles tend to increase distance between cell-cell adhesions, while radial actin bundles arise from those connected to cell-cell complexes. Thus, spatial distribution of traction forces in actomyosin structures seems to be key in permeability of endothelial cells \cite{Oldenburg2014}.

VE-cadherins do not transmit force linearly in cell-cell adhesions. It is a two-phase process. First, in the absence of contraction forces, the predominant adhesions in a monolayer are linear adherens junctions (LAJs). The formation of RSFs in endothelial cells causes movement of the actin cytoskeleton into the cell, which reduces the presence of actin bundles that serve as connecting elements for LAJs. This reduction of actin destabilizes the LAJs and, therefore, for the creation of gap openings in the endothelial wall. Second, this destabilization, caused by contraction forces in the cells, has another opposite effect: the areas with the presence of VE-cadherin adhesions resist, resulting in the appearance of focal adherens junctions (FAJs). It is thought that this process of formation of the FAJs could be a stochastic process or could be induced by unknown inhomogeneities of the LAJs at submicroscopic levels \cite{Oldenburg2014}. When these adhesions begin to withstand traction forces, the $\alpha$-catenin begins to lengthen and vinculin recruitment occurs in the adhesions. Once these bonds have been strengthened, a signaling takes place leading to the restoration of the LAJs. This behavior causes a very active and changing dynamics in the monolayer.

The structure of endothelial cells is determined by the dynamic interaction between cell-ECM adhesions, cytoskeletal networks and cell-cell adhesions whose integrity determines barrier function \cite{Oldenburg2014}. Cell-cell adhesions are formed by a large number of proteins. Among these proteins, VE-Cadherin stands out as a key player in force transmission. Several studies \cite{Lampugnani1995,Taddei2008} have concluded that lost of VE-Cadherin adhesions affects other cell-cell adhesions, deteriorates the integrity of the barrier function and leads to deep changes in the cytoskeletal structure. The VE-Cadherin complex is connected to the actin cytoskeleton through $\alpha$-catenin forming AJs (endothelial adherent junctions) which play a key role in barrier function. Actin bundles close and parallel to cell-cell adhesions improve the integrity of these adhesions, while RSFs in the center of the endothelial cell are associated with lower stability \cite{Lum1996}. Cell-ECM adhesions control the organization and contractility of the actomyosin network. These integrin-based adhesions are organized in FAs (focal adhesions) and constitute signalling centers that sense chemical and mechanical information. Thus, there is a mechanical feedback between the actomyosin cytoskeleton and the ECM through FAs which matches the mechanical environment from cellular to structural level \cite{Ingber2002}.

The aim of this work is to simulate the endothelial monolayer and cell-cell adhesions through a new constitutive model. This model will take into account the dynamic behavior of cell adhesions and its stochastic nature. The model will investigate the influence of blood vessel diameter in the formation of opening gaps in the cellular monolayer through which tumor cells could extravasate. Unlike previous works \cite{Escribano2019} the model will consider the three dimensional (3D) geometry of the vessels and cells, which will be simulated as continuum elements.

\section{Materials and Methods}

\textcolor{black}{In this work, we simulate the endothelial cell monolayer as a continuum medium, where the balance of linear momentum is satisfied in all the monolayer:}
\textcolor{black}{
\begin{equation}
\mathbf{\nabla}\cdot \mathbf{\sigma} + \mathbf{f} = \mathbf{0}
\end{equation}
}

\textcolor{black}{With $\mathbf{\sigma}$ the Cauchy stress tensor and $\mathbf{f}$ the body force per unit of current volume and $\mathbf{\nabla}\cdot$ the divergence operator.}

\textcolor{black}{This equation is solved by means of the Finite Element approach, where we discretized the monolayer domain, distinguishing between the cell body and the cell-cell adhesions. Actually, the cell body is discretised by solid elements, and the cell-cell adhesions are discretized by truss elements that connect nodes from different cell discretizations.}

\textcolor{black}{Therefore, in the following subsection, we present the basics of how cell-cell interactions are simulated. Next, we introduce the mathematical model used to describe the cell mechanical behaviour. Finally, we show how these mathematical models have been numerically implemented in a FE-based approach.}

\subsection*{Modeling and simulations of cell-cell adhesion}
\textcolor{black}{Previous works explain cell-cell adhesion through discrete \cite{Palsson2000}, continuous \cite{ARMSTRONG2006} and hybrid \cite{Gonzalez-Valverde2017} models, they include the effect of cell-cell interactions as interaction forces or potentials, but in general} they not consider explicitly adhesions as a different element.

\textcolor{black}{Experimental evidences show that cadherins bonds increase their lifetime if subjected to mechanical forces \cite{Sokurenko2008,Panorchan2006,Manibog2014,Buckley2014}. Moreover, the catch bond law are widely use in literature to explain cell adhesions in general \cite{Sokurenko2008} and it is a widely accepted model in the modelling literature \cite{Escribano2019,Buckley2014,Manibog2014}. Thus,} the failure of the cell-cell joints is defined as a \textit{catch-slip bond} law \cite{Novikova2013} that provides a stochastic behavior to the rupture of these adhesions. This law adapts to the mechanical behavior observed in VE-cadherins, where the joints subjected to low forces are unstable. As the stress in the joints increases, they become more stable until a point where the joints are not able to withstand the force and the probability of rupture begins to grow exponentially. Thus, we assume the probability of rupture or binding follows this law:

\begin{eqnarray}
	k_{ub}(F)=e^{\varphi_c-\frac{F}{F_{sat}}} + e^{\frac{F}{F_{sat}}-\varphi_s}\\
	prob_{ub} = 1 - e^{-k_{ub} \Delta t}
\end{eqnarray}
where $k_{ub}$ is the ratio of failure which is a function of the ratio between the force $F$ in the cell-cell adhesion and $F_{sat}$ parameter of the union saturation force and $\varphi_c$ and $\varphi_s$ are adimensional force parameters for the curve of \textit{catch} and \textit{slip bond}, respectively. It is assumed that the effect of compressive forces is not able to cause damage to the joints. The probability of rupture, $prob_{ub}$, behaves as an exponential function that depends on the force that the union supports ($k_{ub}(F)$) and the time that the union is supporting this force ($\Delta t$), since in biological processes of rupture, the time that a force acts on a material is crucial in the effect that force has on the material (e.g. pressure ulcers in the skin). We assume the time increment is small enough to consider the force constant during this time interval.

Unbinding is represented by the loss of rigidity of the adhesion by a variable of damage ($d$).  To compute if the adhesion is bound or unbound, we generate a random number ($\tau$) from a uniform distribution between 0 and 1, which is compared with the probability of unbinding:

\begin{equation}
d=\left\{
\begin{array}{lr}
0.00 & \tau < prob_{ub} \\
0.99 & \tau \geq prob_{ub}
\end{array}
\right.
\end{equation}

\begin{figure}[h]
	\centering
	\includegraphics[width=\textwidth]{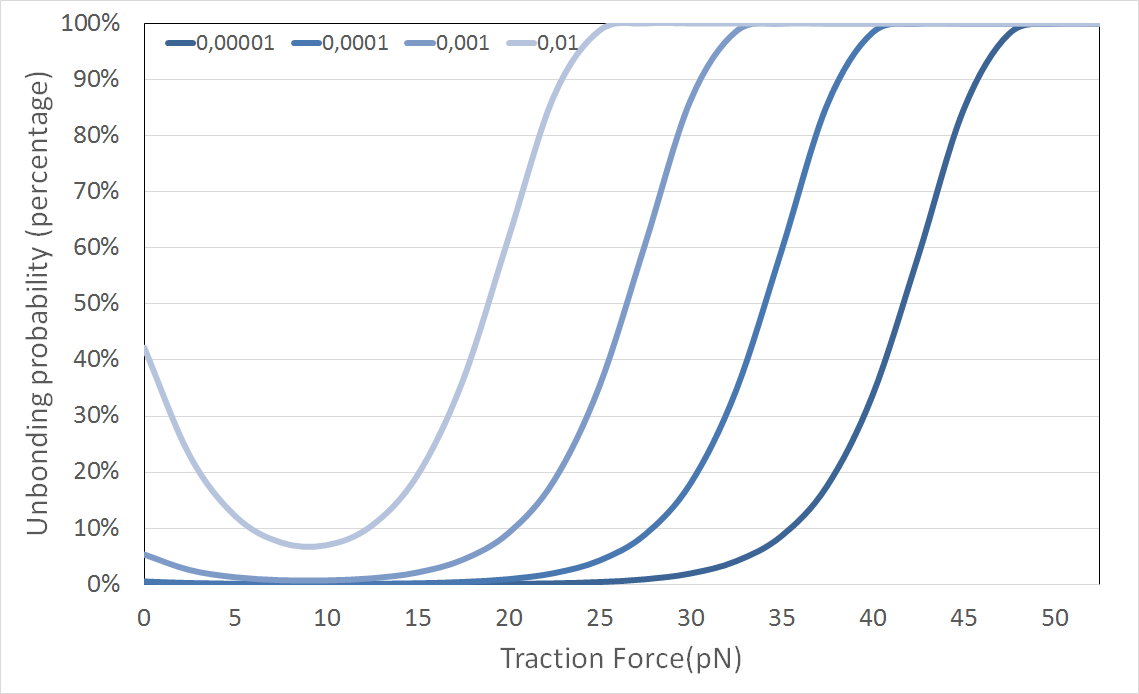}
	\caption{Evolution of the probability of unbinding of cell-cell adhesions as a function of the traction forces they support and their duration ($\Delta t$ in seconds).}
	\label{grafica_catch_bond}
\end{figure}

The evolution of this law (figure \ref{grafica_catch_bond}) follows the behavior of VE-cadherins observed in experiments \textcolor{black}{\cite{Sokurenko2008,Panorchan2006,Manibog2014,Buckley2014}}. When the traction forces are low, there is a high vinculin recruitment by the VE-cadherins, so the FAJs have a lower rigidity, which is reflected in the higher probability of rupture in the area of the graph with negative slope. After this phase, when the traction forces increase and are within a range in the valley of the curve, the joints work in their optimum zone, so they present a low rupture probability. As the traction forces continue to rise, the probability of rupture increases again exponentially. In fact, for the same force, there can be a big difference in the probability of unbinding depending on whether this force is applied in a large ($\Delta t=0.01$ s) or small ($\Delta t=0.00001$ s) time increment.

In a similar way as the law of unbinding, the formation of cell-cell adhesions is also characterized by a stochastic probability of binding ($prob_b$), which depends on the separation between cells:

\begin{eqnarray}
k_b(L_{cadh})=D_{cadh} \left(1-\frac{L_{cadh}}{L_{lim}}\right)\\ \label{union_eq}
prob_b = 1 - e^{-k_b \Delta t}
\end{eqnarray}
where $k_b$ is the binding ratio, $D_{cadh}$ represents the density of VE cadherins available for binding in the monolayer, $L_{cadh}$ is the current length of the adhesion, $L_{lim}$ is the maximum length for which the union is allowed and, $\Delta t$ is the time increment.
 
Binding is represented by the restoration of initial rigidity of the adhesion.  To compute if the adhesion is bound or not we again generate a random number ($\tau$) from a uniform distribution between 0 and 1, which is compared with the probability of binding:

\begin{equation}
d=\left\{
\begin{array}{lr}
0.99 & \tau < prob_{b} \\
0.00 & \tau \geq prob_{b}
\end{array}
\right.
\end{equation}

\begin{figure}[h]
	\centering
	\includegraphics[width=1.0\textwidth]{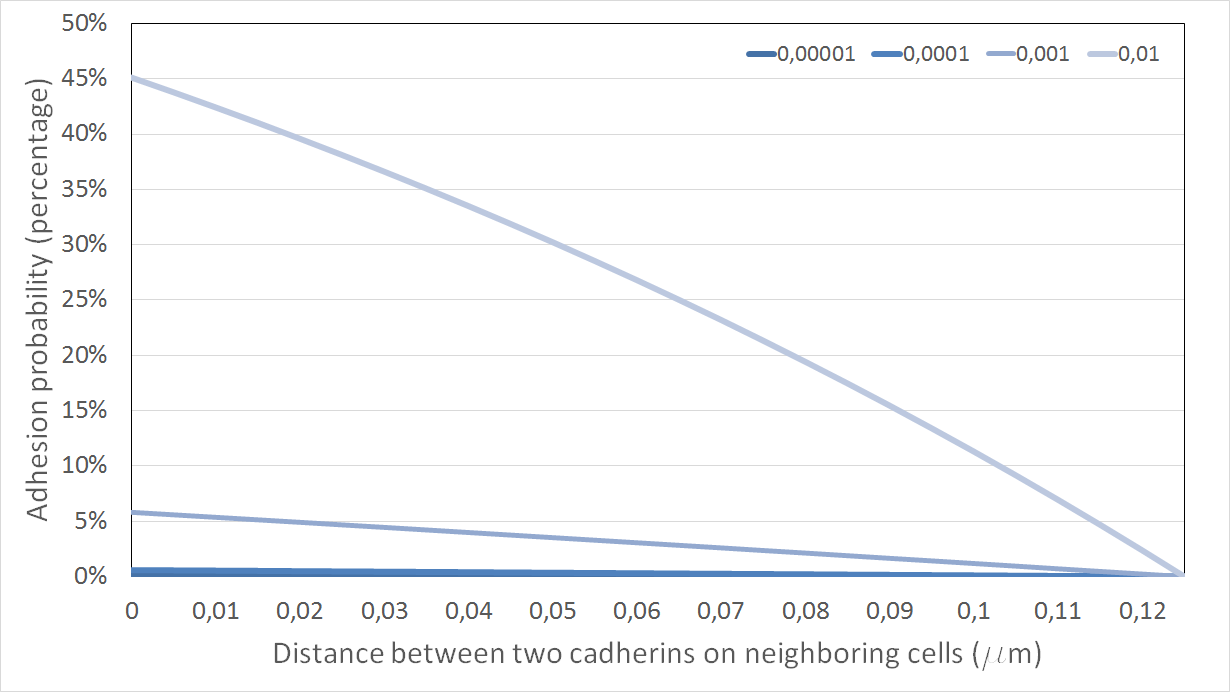}
	\caption{Evolution of the probability of cell-cell adhesions as a function of the distance between two cadherins on neighboring cells and the increase in time (seconds). $D_{cadh}=60$.}
	\label{grafica_union}
\end{figure}

The evolution of the formation of cell-cell adhesions is observed in figure \ref{grafica_union}. Once the maximum cadherin length has been exceeded, the probability that the adhesion occurs is null. The density of VE-cadherin available for union ($D_{cadh}$) modifies the slope of the function, so it establishes a maximum union probability of 45\%.

When the adhesion is unbound we assume that the mechanical properties of the cadherin is reduced to 1\% of its original value due to its damage. In contrast when it binds again, adhesion recovers its initial mechanical properties.

\subsection*{Modeling cell mechanics}

\textcolor{black}{We consider cells behave as an hyperelastic Neo-hookean material. Thus, the strain energy function reads:}

\textcolor{black}{
\begin{equation}
U=C_{10} (\bar{I}_{1}-3)+\frac{1}{D_{1}}(J-1)^{2}
\end{equation}
Where $\bar{I}_{1}$ is the first invariant of the right Cauchy-Green deformation tensor, $J$ is the elastic volume ratio and $C_{10}$ and $D_1$ are material parameters. For endothelial cells, material parameters are set to 173Pa and $6\cdot10^{-4} Pa^{-1}$ respectively which are equivalent to an initial elastic modulus of 1000Pa and a Poisson ratio of 0.45 based on experimental data from previous works \cite{Zeng2010,Guz2014,Alcaraz2003}.
}

We assume finite deformations due to the large strains that VE-cadherin suffers during gap formation.

In the case of blood vessels of infinite radius compared to the endothelial cell, we will adopt a plane stress assumption. \textcolor{black}{Nevertheless}, for the blood vessel of known diameter a 3D model is adopted (figure \ref{geometrias}).

The endothelial monolayer is a very dynamic structure exposed to very variable boundary conditions. Thus, certain simplifying hypotheses have to be established in order to carry out its modeling. The endothelial cells are immersed in an ECM, which acts as a substrate that holds them in place when mechanical conditions are not severe, and they are also exposed to the internal pressure exerted by blood flow which in this first model is neglected. \textcolor{black}{Henceforth}, the radial displacement experienced by the monolayer can be assumed to be relatively small and have little influence on the rupture of cell-cell adhesions. We assume endothelial cells are perfectly anchored to the vessel wall through the basal membrane which allow displacements in the circumferential and longitudinal direction and avoid radial displacements. Therefore, we neglect the deformation of the blood vessel diameter. Hence, in the 3D endothelial monolayer model displacements in the radial direction are set to zero. In addition, at the free ends of the monolayer, displacements are constrained in all directions,we assume the boold vessel is long enough, thus this boundary condition do not affect model results.

The cell monolayer is continuously subjected to normal and tangential stresses caused by the blood flow circulating within it. However, these stresses are not directly causing changes in endothelial permeability. One of the most significant permeability mechanisms of the endothelial monolayer is through increased intracellular Calcium concentration ($Ca^{2+}$). The increase in Calcium concentration activates signaling pathways that affect both the structural organization of the cytoskeleton and the cell-cadherin-VE adhesions \cite{Tiruppathi2002}. This is because Calcium-dependent proteins cause the endothelial cells to contract, leading to increased traction forces supported by adhesions. The propagation of Calcium through waves has been widely accepted as a factor which changes the architecture of the endothelial monolayer \cite{Junkin2013}.

\begin{figure}[h]
	\centering
\begin{subfigure}{0.65\textwidth}
	\includegraphics[width=\textwidth]{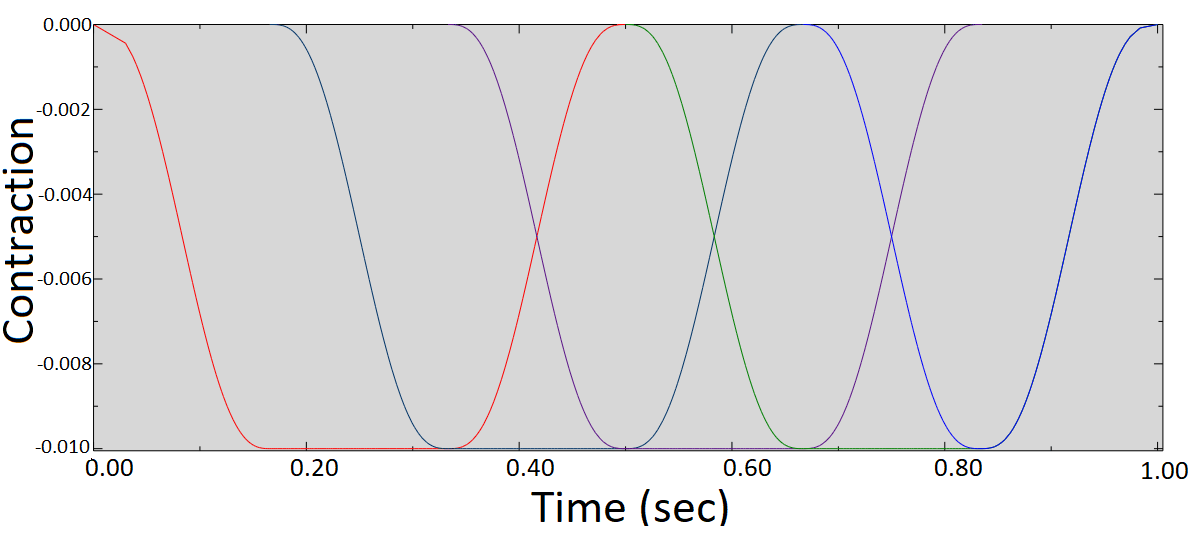}
    \caption{}
\end{subfigure}
\begin{subfigure}{0.15\textwidth}
     \includegraphics[trim={9cm 0  0  0},clip,width=\textwidth]{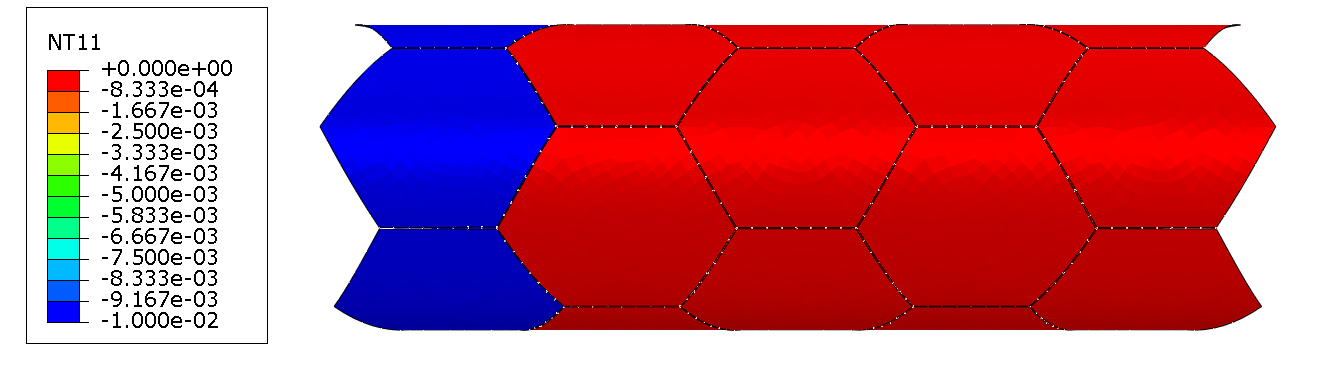} \caption*{\textcolor{black}{t=0.250sec}}
       \includegraphics[trim={9cm 0  0  0},clip,width=\textwidth]{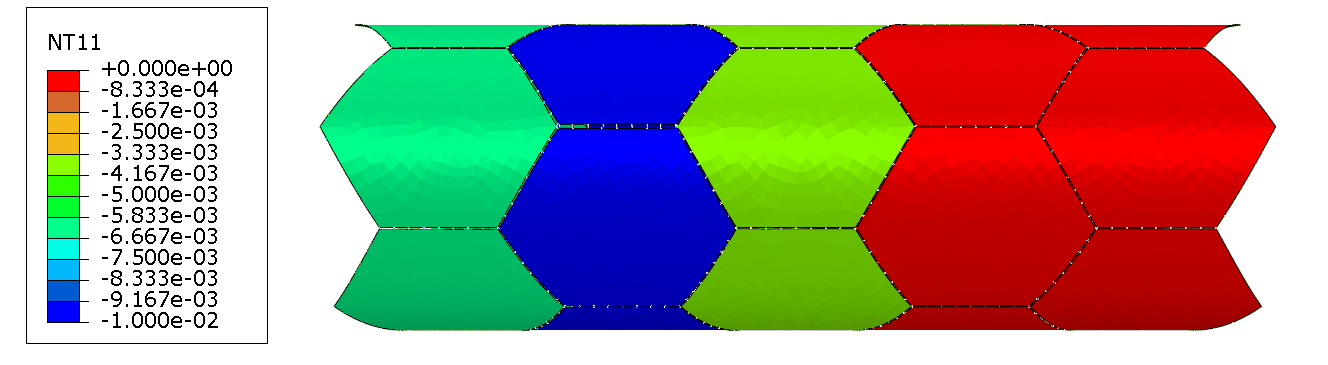}\caption*{\textcolor{black}{t=0.416sec}}
         \includegraphics[trim={9cm 0  0  0},clip,width=\textwidth]{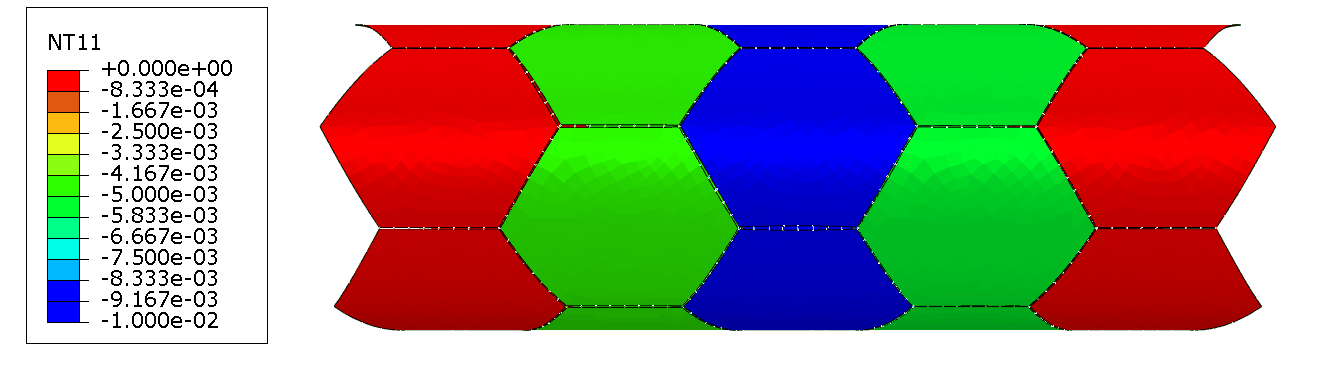}\caption*{\textcolor{black}{t=0.583sec}}
           \includegraphics[trim={9cm 0  0  0},clip,width=\textwidth]{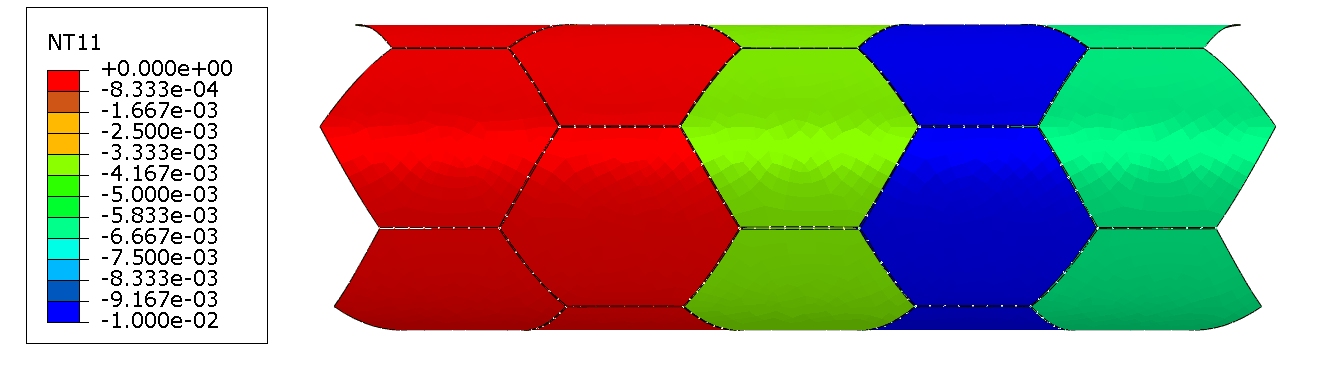}\caption*{\textcolor{black}{t=0.750sec}}
       \caption{}
\end{subfigure}
\caption{Evolution of the contraction due to calcium wave for each column of cells as a function of time, \textcolor{black}{red lines correspond to the first cell column, dark blue for the second column, purple to the third cell column, green for the fourth and so on} (a), and spatial evolution of this contraction \textcolor{black}{at different timepoints} in the blood vessel, red means minimum contraction and blue maximum, green means intermediate value of contraction (b).}
\label{ola_temperatura}
\end{figure}

The loading conditions simulated in the model are these Calcium waves in order to observe the effect they have on the rupture and remodelling of the VE-cadherins and, therefore, on the permeability of the endothelial monolayer. The wave begins at one end of the endothelial cell monolayer and advances in the longitudinal direction of the geometry. The distribution of the contraction is shown in figure \ref{ola_temperatura} where each  line represents a different column of cells, its amplitude is estimated between 1\% and 3\% \cite{Sun2016}.

To simulate cell contraction due to calcium wave propagation, we assume this calcium wave produces volumetric cell contraction or expansion. \textcolor{black}{At any time we consider three configurations, the undeformed configuration ($\Omega_{0}$), the deformed configuration ($\Omega_t$) and an intermediate configuration of contraction (or expansion) due to the calcium wave ($\Omega_{Ca+}$) which in general is not compatible      \cite{GARIKIPATI2004,ReinaRomo2010}}.

\textcolor{black}{The total deformation gradients maps any point in the undeformed (material) configuration ($\mathbf{X}$) to a point in the deformed (spatial) configuration ($\mathbf{x}$):}

\textcolor{black}{
\begin{equation}
\textbf{F}=\frac{\partial \mathbf{x}}{\partial \mathbf{X}}
\end{equation}
}

We make use of the multiplicative decomposition \cite{Vujosevic2002} of the total deformation gradient $\textbf{F}$:

\begin{equation}
\textbf{F}=\textbf{F}_{e}\cdot\textbf{F}_{Ca+}
\end{equation}
where $\textbf{F}_{e}$ is the isothermal deformation gradient and $\textbf{F}_{Ca+}$ is the deformation gradient produced by the volume change due to the calcium wave:

\textcolor{black}{
\begin{equation}
\textbf{F}_{Ca+}=-a_i(t)\alpha c_{Ca+}\textbf{1}
\end{equation}
}
where $c_{Ca+}$ is the calcium concentration, $\alpha$ is a constant, $\textbf{1}$ is the second order unit tensor and \textcolor{black}{$a_i(t)$ is evolution in time of the amplitude of the calcium wave for each cell column ($i$)}.

Note that these equations are mathematically similar to equations describing thermoelastic systems, \textcolor{black}{to see the full formulation refer to} \cite{Vujosevic2002}.

\textcolor{black}{This calcium wave produces contraction in the endothelial monolayer and it is assumed to be moving through the different cell columns with time. As a first approach, the amplitude of the calcium wave in each cell column ($i$) is assumed to depend just on time (t). The calcium wave enters smoothly in the cell column, it remains for a sixth of a second and it also leaves the cell column smoothly. Thus, the amplitude ($a_i(t)$) of the calcium wave in each cell column is described through:}

\textcolor{black}{
\begin{equation}
a_i(t)=\left\{
\begin{array}{lr}
0.0 & 0 \leq t \leq t_{i} \\
\xi^3(10-15\xi+6\xi^2) & t_{i} \leq t \leq t_{i+1} \\
1 &  t_{i+1} \leq t \leq t_{i+2} \\
1-\xi^3(10-15\xi+6\xi^2) & t_{i+2} \leq t \leq t_{i+3} \\
0 & t_{i+3} \leq t 
\end{array}
\right.
\end{equation}
where $t=\frac{t-t_{i}}{t_{i+1}-t_{i}}$ and $t_i=\frac{1}{6}i$.
}

Due to the small dimensions of the adhesions with respect to the cell size, volume change of the adhesions due to calcium wave is neglected.

\begin{figure}[h]
	\centering
	\includegraphics[width=1.0\textwidth]{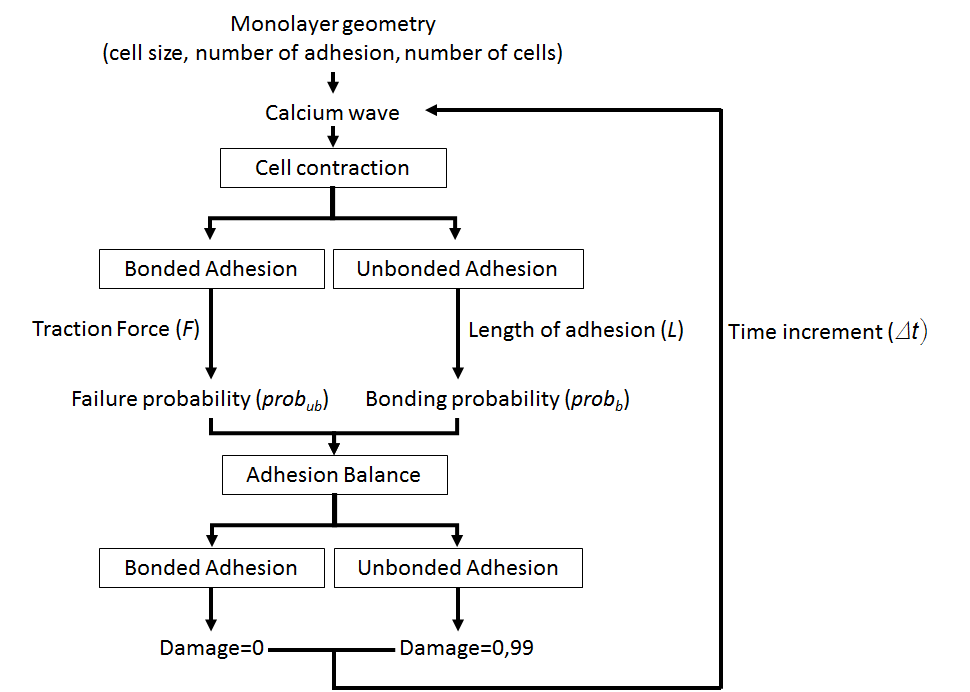}
	\caption{Scheme of the model, once we fix the geometry of the monolayer the calcium wave contracts the cells modifying the mechanical state of the cells and the adhesions. When an adhesion is bonded it can unbond depending on the force in the adhesion and the time increment; when an adhesion is unbonded it can rebond if it reduces its length. After balance of adhesions, mechanical properties of adhesions are updated and a new time increment starts.}
	\label{Modelscheme}
\end{figure}

In summary, the model simulates the binding and unbinding of cell-cell adhesions depending on the mechanical environment they are subjected to, their deformation (length of the adhesion with respect to the maximum) and traction forces transmitted through the adhesion. The only load considered is the calcium wave which contracts the cells and deforms the adhesions (Figure \ref{Modelscheme}).

\subsection*{Numerical Implementation}
To carry out the simulations, we use the commercial FE software ABAQUS. We develop a material user subroutine (UMAT) to define the mechanical behavior of the truss elements that simulate cell-cell adhesion and  several Python scripts to create the geometry of the model. \textcolor{black}{We assume cells are arranged in the monoloyer as regular hexagons following previous works \cite{Escribano2019,Schmedt2012,ODEA2012,Ye2014}}. In fact, to generate the geometry we fix four main parameters: the characteristic length of the endothelial cells (in $\mu$m), which corresponds to the length of the side of the hexagonal cell, the number of VE-cadherins in each side of the cell and the number of endothelial cells that form the endothelial monolayer and the diameter of the vessel. \textcolor{black}{To discretize the model, in the case of the cells, we use linear shell element (S3, S4) for the 3D vessels  and plane stress quadrilateral or triangular linear elements (CPS3, CPS4) in the vessels of very large radious; whereas for the adhesions we use linear truss elements (T3D2 in 3D and T2D2 in 2D).}

\begin{figure}[h]
	\centering
		\includegraphics[width=0.8\textwidth]{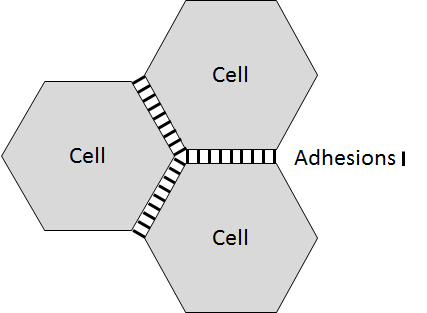}
	\caption{Scheme of cell geometry generation and cadherins attachment points}\label{sketch_cell}
\end{figure}

Next, we present all the step that we use to define the FE model to simulate endothelial cell monolayers:

\begin{enumerate}
\item We generate a cell. All the cells of the monolayer will be generated from this first one, we assume homogeneity of geometry and structure between cells. We start by generating the hexagonal cell with side equal to the value provided by the user (figure \ref{sketch_cell}). We estimate a thickness of 500 nm for the endothelial monolayer \cite{Pries2000,Feletou2011}, however there is high variability for the thickness in literature \cite{Stroka2011}.
\item To create cell-cell adhesion, we partition cells sides, the algorithm created will iterate $n$ times making a number of partitions such that the number of adhesions per side is $2^n + 1$ (figure \ref{sketch_cell}, with $n=3$).
\item With the cell already defined, the structure of the endothelial monolayer is formed. Cells are positioned at a fixed distance defined by the user, which corresponds to the length of the VE-cadherins (50 nm for our simulation \cite{Davies1993}). The cells are transferred symmetrically in such a way to obtain the distribution described by the user based on the number of cells in each column.
\item After obtaining the geometric distribution of the cells in the undeformed configuration, we run a \textit{script} which can determine the position of FAJs in the monolayer. First, we identify all points of interest likely to form a cell-cell adhesion. Second, this list of points is compared to the relative distance between subsets of two points to know if they form an adhesion. We verify whether the subset of points is at a distance equal to the length of the VE-cadherins. If it is and the points belong to two different cells, a cell-cell adhesion is created. The initial model is two dimensional and a change to cylindrical coordinate system transforms it into a three dimensional monolayer.
\end{enumerate}


We assume VE-cadherins support just axially load and they follow the catch bond and rupture law previously described. Therefore, FAJs work only axial, so working with elements that also withstand bending stresses (e.g. \textit{beams}) would not adequately simulate their behavior; if the cells joined by an adhesion suffer opposite parallel displacements (shear stresses), the elements would be acting as LAJs (tangential stresses) since their arrangement would be nearly parallel to the membrane.

Due to the dynamic behavior of VE-cadherins, these may suffer many changes, binding and unbinding, during the simulation. When the adhesion is defined as unbound its elastic modulus is reduced to 1\% of the original value, thus there will be a loss of stiffness in the joint. These losses of stiffness cause a great non-linearity in the stresses suffered by the adhesions making it difficult to converge in a static analysis, since FE program will continue to iterate until these discontinuities are small enough and the tolerances of the equilibrium are satisfied. 


Thus, non-linear static analyzes can be unstable. In our case, changes in material stiffness can lead to local instabilities where there is a transfer of strain energy from one part of the model to another, causing the global methods used to arrive at the solution that does not work. This is solved by automatic stabilization with a constant damping coefficient, in case a region becomes unstable, it absorbs part of the deformation energy dissipating it through the damping. \textcolor{black}{In fact, the numerical solution of cancer related simulations is complex \cite{DEHGHAN2018,MOHAMMADI2019}.}

\textcolor{black}{All the simulations were executed on a High-Throughput Computing (HTC) environment which can reduce the required times to run those simulations by parallelizing them. The averaged execution time of each simulation was 5 hours using 1 CPU and 1024 MB of RAM.}

The VE cadherins suffer large deformations, so evaluating the deformations and stresses with respect to their undeformed configuration would not provide real solutions. Therefore, the simulation is performed with the hypothesis of finite strains. The model parameters used in the simulation are summarized in Table \ref{ModelParameters}.

\begin{figure}[h]
	\centering
	\begin{subfigure}{0.35\textwidth}
		\includegraphics[width=\textwidth]{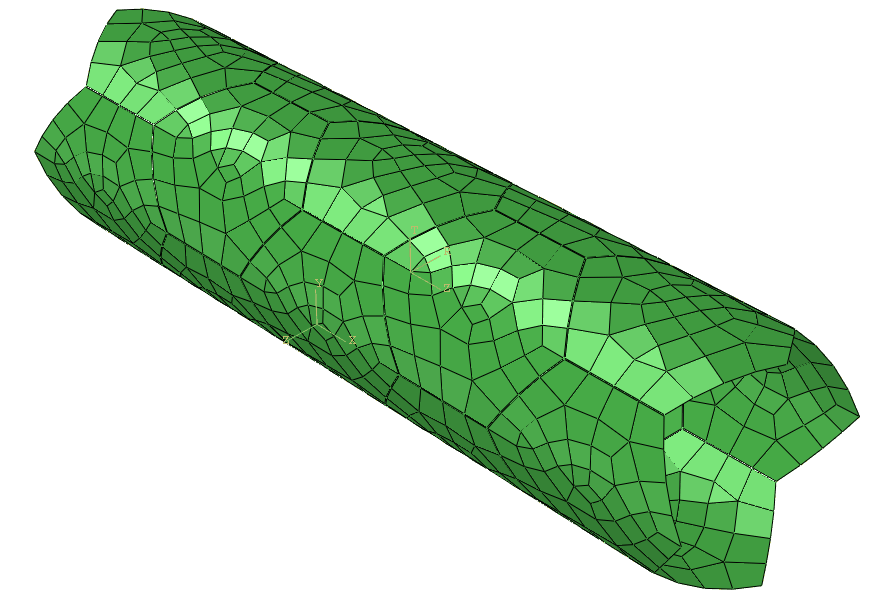}
        \caption{}
		\label{3D_small}
	\end{subfigure}
\begin{subfigure}{0.35\textwidth}
		\includegraphics[width=\textwidth]{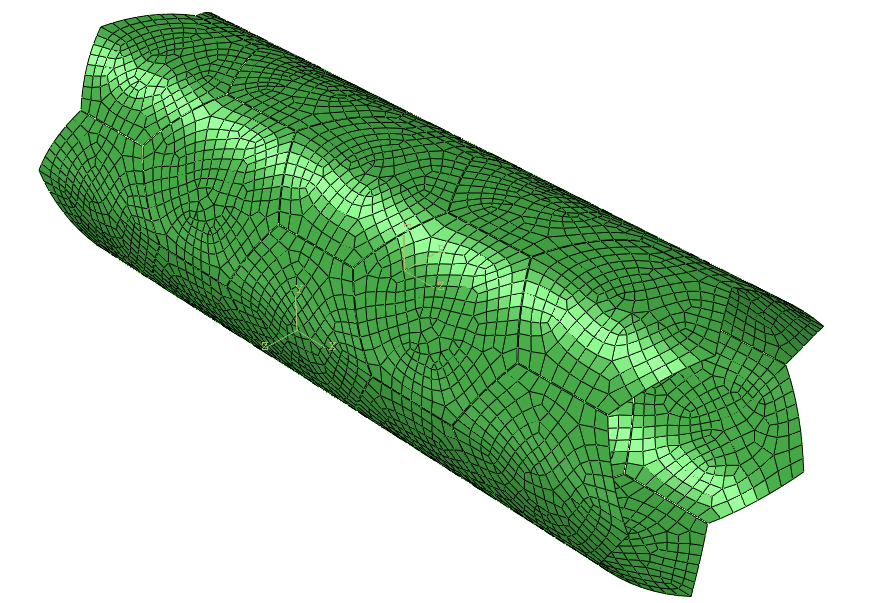}
        \caption{}
		\label{3D_normal}
	\end{subfigure}
	\begin{subfigure}{0.35\textwidth}
		\includegraphics[width=\textwidth]{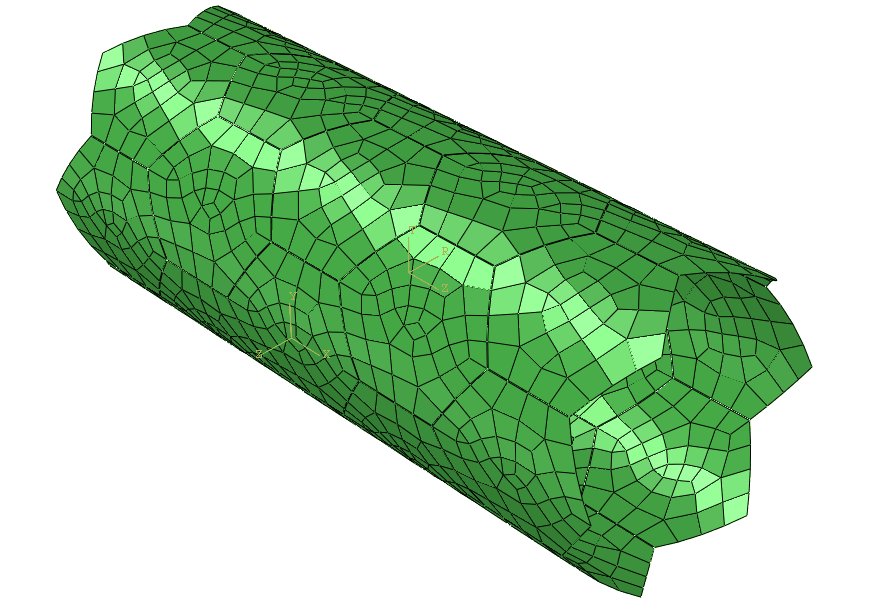}
        \caption{}
		\label{3D_big}
	\end{subfigure}
\begin{subfigure}{0.35\textwidth}
		\includegraphics[width=\textwidth]{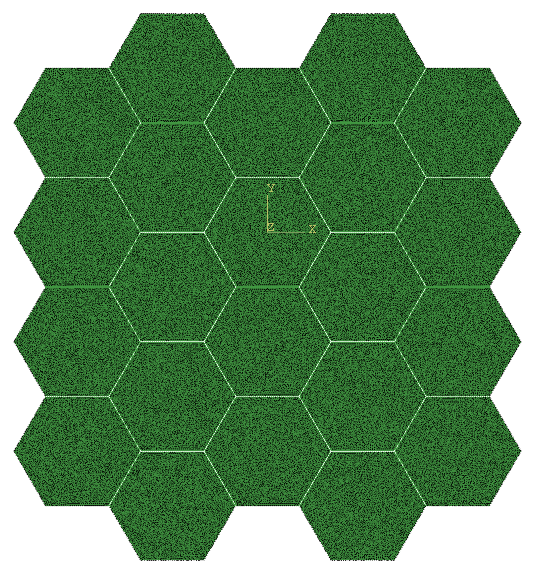}
         \caption{}
		\label{3D_infinity}
	\end{subfigure}
	\caption{Three-dimensional models studied: (a) small diameter ($10.1\mu$m), (b) medium diameter ($13.86\mu$m),(c) large diameter($16.64\mu$m) and (d) very large diameter ($\infty$)}
\label{geometrias}
\end{figure}

Diameters of blood vessels are highly variable. In the bat's wing these diameters range from 76.2$\mu$m and 52.6$\mu$m in the veins and arteries respectively to 3.7$\mu$m in a capillary \cite{Wiedeman1963}. \textcolor{black}{However, in humans the diameter of blood vessels ranges from around 8$\mu$m in capillaries to more than 1cm in large arteries \cite{Aird2005,Traore2017}.} In this work, we study three different blood vessel diameters 11.1$\mu$m, 13.86$\mu$m, 16.64$\mu$m  and an infinite diameter of the blood vessel compare to the cell diameter (10$\mu$m) (figure \ref{geometrias}). To determine if the blood vessel diameter has an influence in the formation of openings in the blood vessels.

\textcolor{black}{In simulations, we use models with a maximum of 33 adhesions per cell side which show a good compromise between accuracy and computational cost. If we increase the number of  cadherins per cell side, results would be more realistic; however, it would considerably increase the computation time to calculate the cell-cell adhesions.}

\begin{table}[]
\centering
\begin{tabular}{|l|l|}
\hline
$\varphi_c$ & 2.5          \\ \hline
$\varphi_s$ & 0.2          \\ \hline
$F_{sat}$    & 3.3$pN$      \\ \hline
$L_{lim}$  & 0.125$\mu m$ \\ \hline
$D_{cadh}$ & 60 \\ \hline
\end{tabular}
\caption{Model parameters used in the simulation.}
\label{ModelParameters}
\end{table}

We simulate the mechanical behavior of the endothelial monolayer and the damage suffered by the cell-cell adhesions caused by a Calcium wave that travels through the monolayer. The intensity of the cellular contraction has been varied for three different cases ($\alpha c_{Ca+}= 0.01, 0.02, 0.03$).

\section{Results and Discussion}
\begin{figure}[h]
	\centering
	\begin{subfigure}{0.8\textwidth}
		\includegraphics[width=\textwidth]{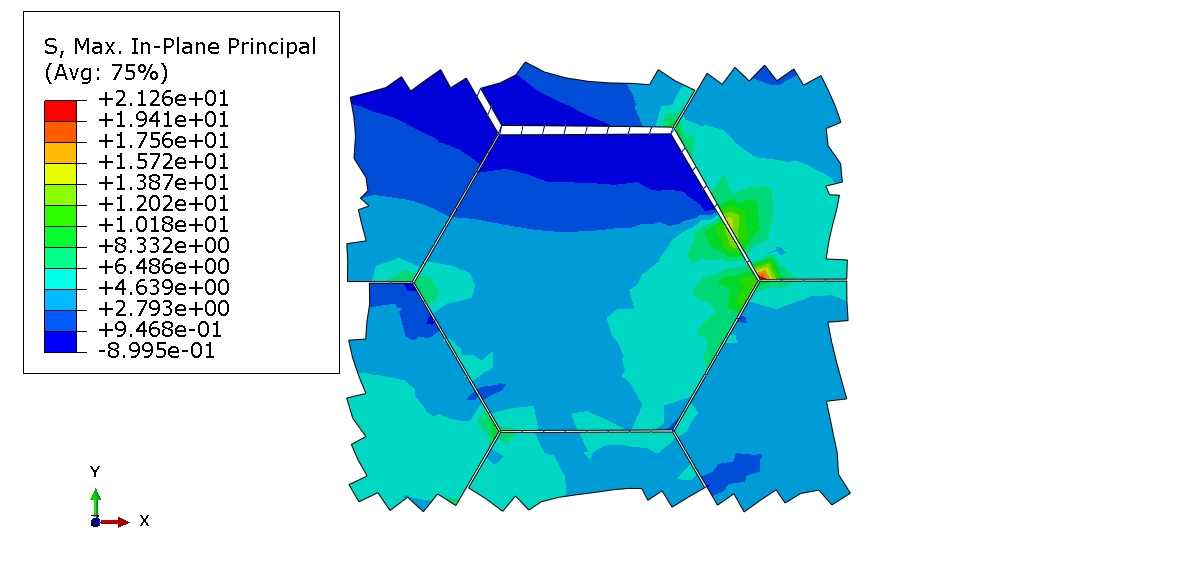}
		\caption{2D model (Stress in $Pa$)}
		\label{2D_concentracion}
	\end{subfigure}
	\begin{subfigure}{0.8\textwidth}
		\includegraphics[width=\textwidth]{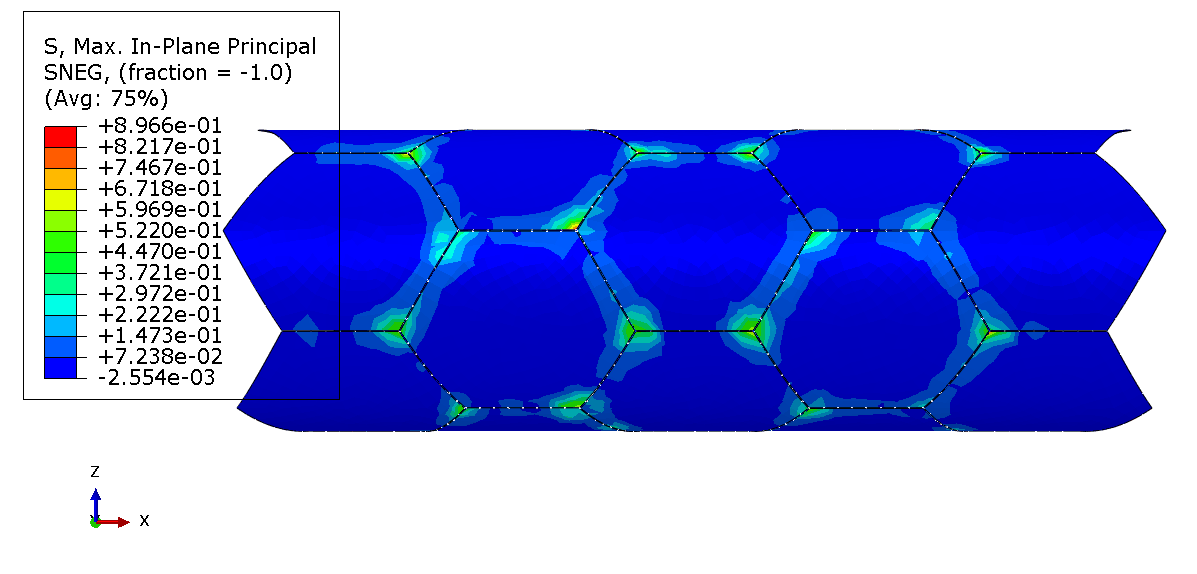}
		\caption{3D model (Stress in $Pa$)}
		\label{3D_concentracion}
	\end{subfigure}
	\caption{Stress concentration in the Endothelial Monolayer (Pa).}\label{concentraciones}
\end{figure}

First, we analyse the location of the stress concentrations in the endothelial cells. The maximum stresses are generally found in the vertices where three different cells meet (Figure \ref{concentraciones}).

These results indicate that a higher stress in the joints located at these points made them more prone to break as they withstand higher stresses than others (Figure \ref{concentraciones}). This implies that the openings in the endothelial monolayer are normally formed in the adhesions located in these vertices, suggesting that extravasation is more likely to occur in these areas.

\begin{figure}[h]
	\centering
	\begin{subfigure}{0.25\textwidth}
		\includegraphics[width=\textwidth]{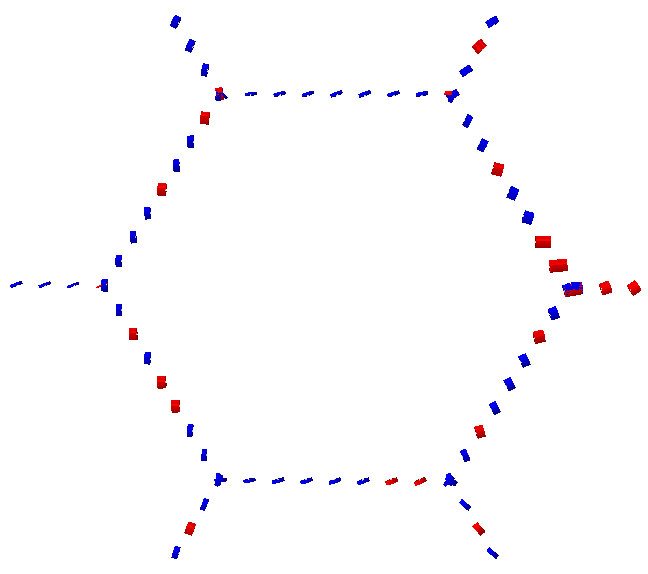}
		\caption{}
		\label{2D_inestabilidad}
	\end{subfigure}
	\begin{subfigure}{0.475\textwidth}
		\includegraphics[width=\textwidth]{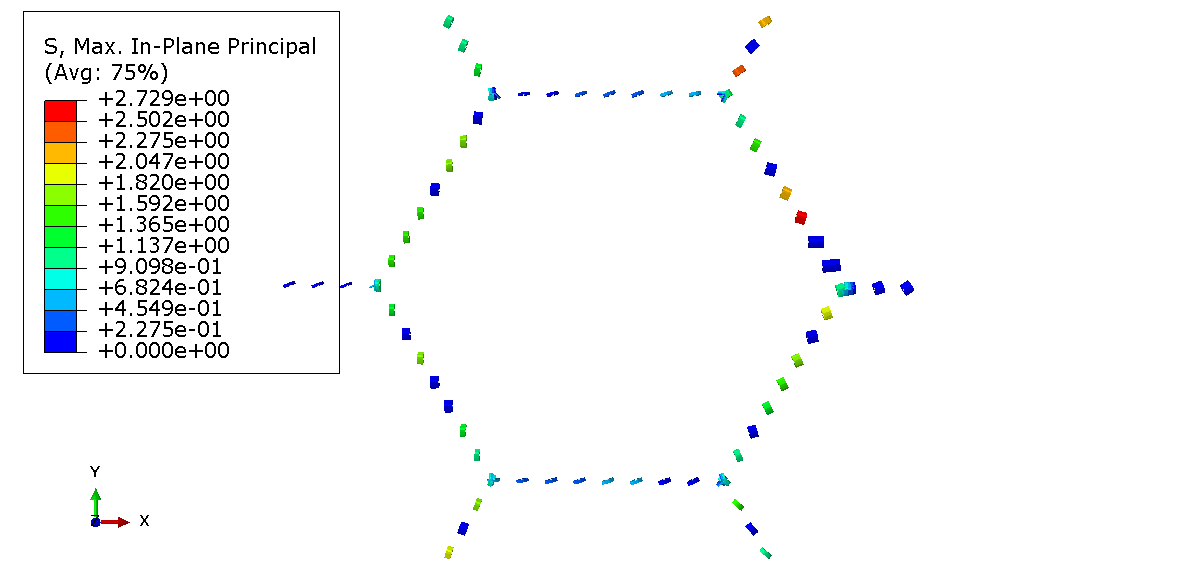}
		\caption{}
		\label{3D_inestabilidad}
	\end{subfigure}
	\caption{Unstable cell-cell adhesions in the endothelial monolayer. Only the joints are represented, as the cell in this case does not provide any information: (a) state of the adhesion: bound in blue and unbound in red; (b) Maximum principal stress in \textcolor{black}{cell-cell adhesions} ($Pa$). The appearance of the elements has been modified to improve visibility.}\label{inestabilidad}
\end{figure}

We clearly observe different failure patterns of the endothelial monolayer. The endothelial monolayer is a very active and changing dynamic structure. This causes its mechanical behavior to be very different for different mechanical environments. Three main phases of work can be distinguished for VE-cadherins: the resting phase, the working phase and the rupture phase. The resting phase is unstable, this situation occurs when the contraction forces in a monolayer are very low, that is, the Calcium wave circulating in the monolayer does not cause a high contraction, so the stresses experienced by VE-cadherins are very low (Figure \ref{inestabilidad}). When the traction forces are very low there is no vinculin binding recruitment \cite{Kruse2015}, thus the unions are not very stable and their rigidity is low. This situation is reflected in the model, the monolayer presents a high damage in the VE-cadherins. However, since there is very little deformation of the adhesions, the possibility of their recovery is  high (equation \ref{union_eq}), which creates a very active dynamic behaviour in which the unions unbind and bind constantly. This phase would correspond to the first part of the figure  \ref{grafica_catch_bond} where the probability of rupture is high just before reaching the ``valley'' where this probability decreases.

\begin{figure}[h]
	\centering
	\begin{subfigure}{0.5\textwidth}
		\includegraphics[width=\textwidth]{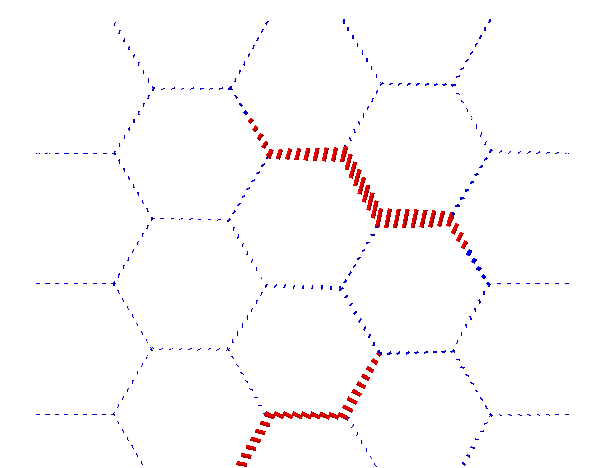}
		\caption{}
		\label{grieta_monocapa_2D}
	\end{subfigure}
	\begin{subfigure}{0.5\textwidth}
		\includegraphics[width=\textwidth]{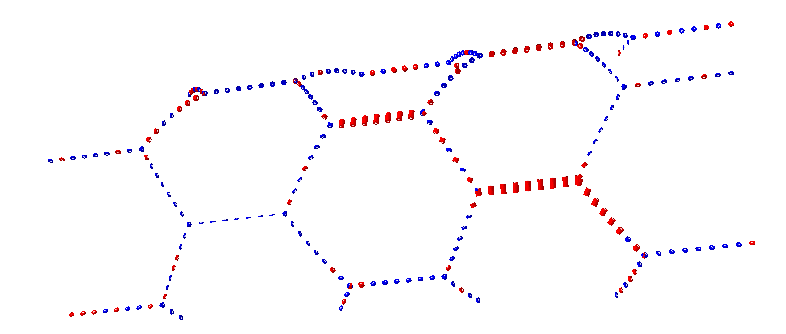}
		\caption{}
		\label{grieta_monocapa_3D}
	\end{subfigure}
	\caption{Pattern of rupture of the endothelial monolayer: (a) 2D model, blood vessel of radius infinity; (b) 3D model blood vessel of small radius. Blue means bound adhesion and red unbound adhesion.}\label{grieta_monocapa}
\end{figure}

\begin{figure}[h]
	\centering
	\includegraphics[width=0.5\textwidth]{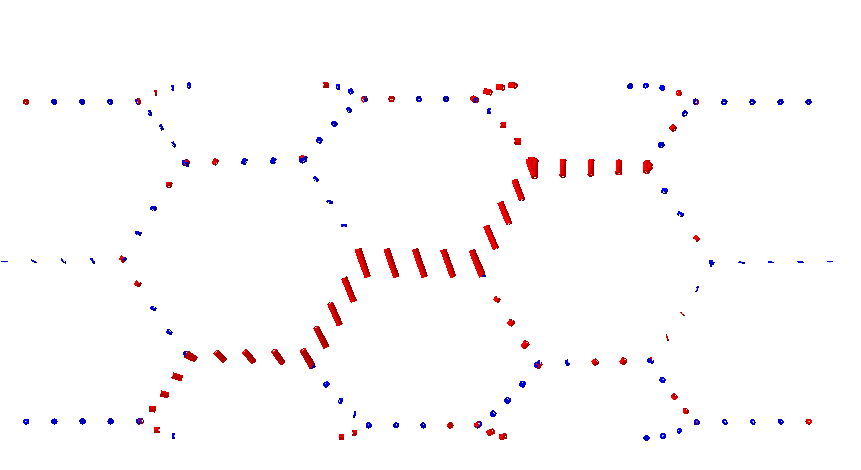}
	\caption{A crack generated in an endothelial monolayer with a radius of $8.32\mu m$ when there is a cellular contraction caused by a Calcium wave. Blue means bound adhesion and red unbound adhesion.}
	\label{grieta_monocapa_3D_grande}
\end{figure}

When the forces exceed the failure limit of cell-cell adhesions, they begin to break. The more adhesions are broken, the more stresses must withstand the remaining adhesion so their probability of unbinding increases. This made the openings in the monolayer to be similar in shape to the propagation of a crack (figure \ref{grieta_monocapa}). This does not occur in normal working conditions of the endothelial monolayer, but it could happen in extreme cases in which the integrity of the monolayer is compromised. For the three-dimensional studied cases, the greater the severity of the failure, the greater the radius of the endothelial monolayer (figure \ref{grieta_monocapa_3D_grande}). In larger monolayers there is a greater effect of the contraction on the forces suffered by the adhesions as the deformation increases, which causes greater damage higher unbinding to the integrity of the endothelial barrier.

After the breaking phase, once the Calcium wave has passed, the cellular contraction decreases and the joints are restored depending on whether they recover their original geometry. In the binding process, a ``zip'' effect is created where broken adhesions begin to bind mostly once the cellular contraction ceases. This behavior is not the behavior that usually occurs in a monolayer, but it would be an extreme case of rupture. A more realistic scenario would be when the monolayer withstands stochastic damage resulting in the generation of specific gaps in certain areas between the cells.

\begin{figure}[h]
	\centering
	\includegraphics[width=0.9\textwidth]{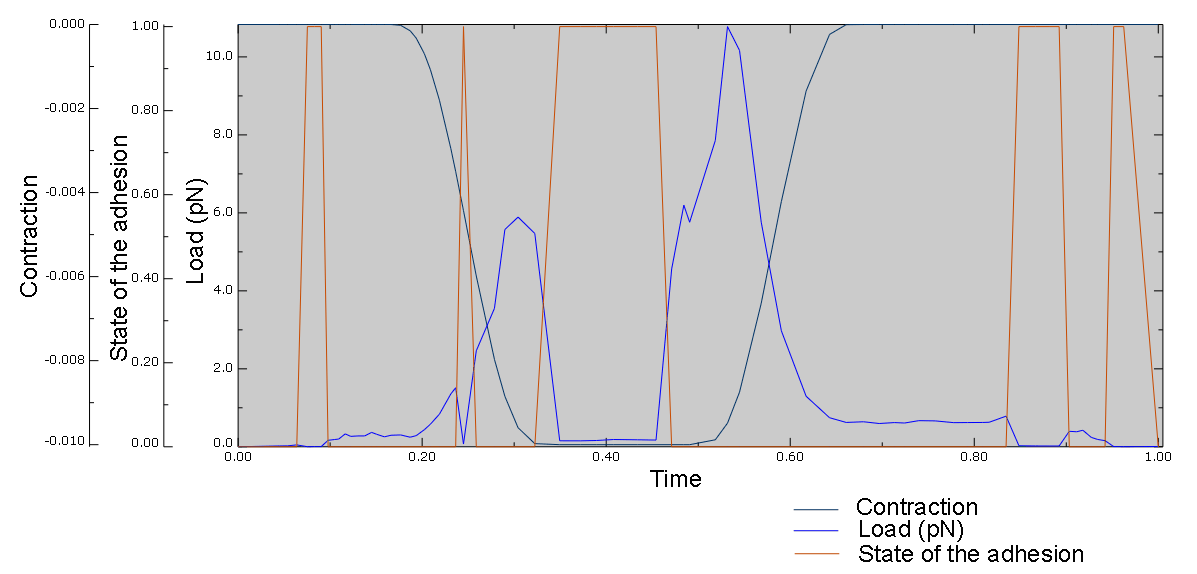}
	\caption{Diagram of evolution of load and state, bound (1) or unbound (0), of a cell-cell adhesion as a function of time \textcolor{black}{(in seconds)}. The shown contraction is that the cells suffer when in contact with the adhesion.}
	\label{grafica_S_NDV1_DELTAT}
\end{figure}

In figure \ref{grafica_S_NDV1_DELTAT} we can observe the behavior of a single VE-cadherin connection. At the beginning, in the absence of contraction, the force supported by the adhesion is almost zero and it unbinds because of the instability characteristic of this phase (left part of figure \ref{grafica_catch_bond}). When the contraction begins to increase and the tensile forces increase, a rupture is observed (in $F=6 pN$). As there are no very high deformation, the adhesion is restored at around $0.45$ seconds and can resist forces higher than $10 pN$ before the contraction disappears. When it is nearly free of forces in the final phase, it returns to the initial phase of instability with two unbinding events at low forces.

Finally, to investigate the influence of the radius of the blood vessel in the appearance of gaps we define the average damage of the adhesion in a period of time ($T$) as:

\begin{equation}
\overline{d}_T=\frac{\sum\limits_{j=1}^{m} \sum\limits_{j=1}^n d_{ij}}{n m}
\end{equation}
where $n$ is the number of increments in the analysis and $m$ the number of total adhesions in the model.

\begin{figure}[h]
	\centering
	\includegraphics[width=0.9\textwidth]{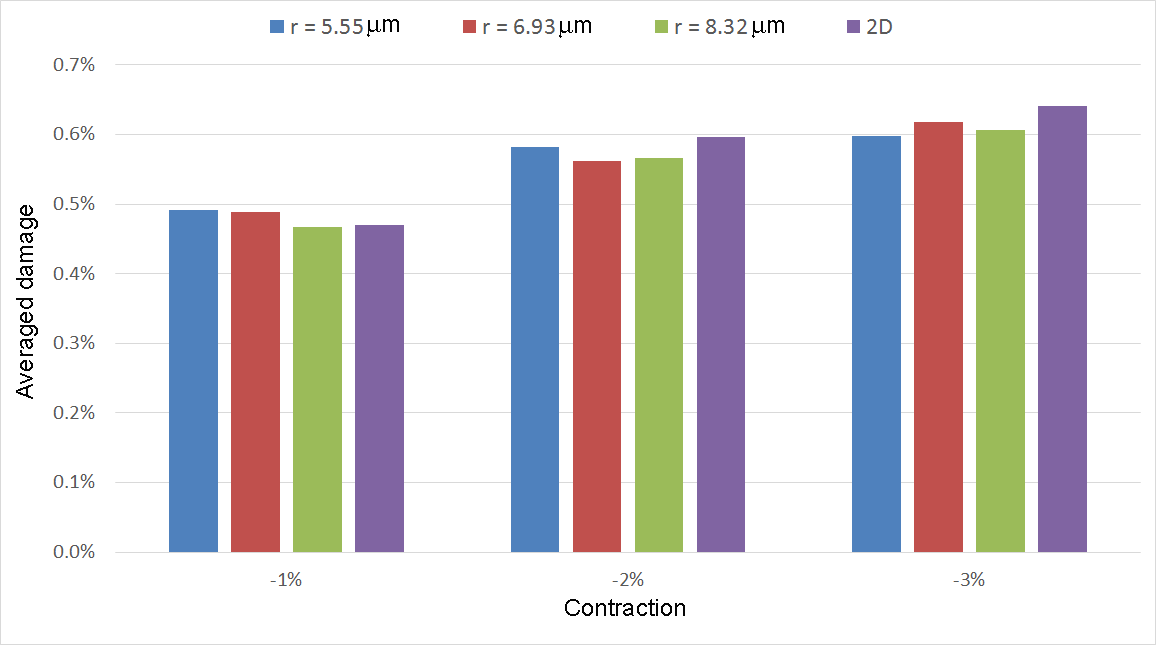}
	\caption{Averaged damage in the adhesion for the different blood vessels radius.}
	\label{averagedamage}
\end{figure}

Average damage in the adhesions increases when the contraction of cells increases. For the smaller radius it increases from 0.49\% for a contraction of 1\% to 0.64\% for a contraction of 3\%. However, there are no significant changes when the radius of the blood vessel increases, for example, for a contraction of 2\%, we found that average damage ranges from 0.57\% to 0.595\% for all simulated blood vessel radius.

\section{Conclusions}

In this work, we have proposed a hybrid model of three-dimensional endothelial monolayers. This model considers cells as linear elements of small thickness and adhesions as truss elements that can be bound and unbound depending on the mechanical environment by means of a catch-bond law. There are several models in literature, which study adhesion and gap formation in blood vessels \cite{Escribano2019}. However, as far as we know, this is the first work, which simulates gap formation in three dimensions considering the blood vessel curvature. Even though we have observed no significant effect of the blood vessel diameter on the rupture of the adhesions of monolayer. The model shows how stress is more likely to accumulate in the vertices between three cells targeting this areas as location where extravasation is more likely to occur. This fact is in agreement with observation in other previous works \cite{Jeon2013,Escribano2019} where they observed in \emph{in-vitro} experiments that gaps are more likely to occur in these vertices than in a two cell border.

To develop these simulations, several simplifications were necessary. First, we assume that the blood vessel diameter is constant during the simulation, so it does not change in response to cell contraction. Thus, we assume endothelial cells are perfectly joined to the vessel wall in the radial direction, and the vessel wall is rigid. A more realistic approach should include the properties of the vessel and the characteristics of the adhesion of endothelial cells to the cell vessels, however deformation of cells in the radial direction will be smaller than the circumferential or longitudinal one. Second, adhesions could be activated when bound or deactivated when unbound; nevertheless, no new adhesions could be created when they were not present at the beginning of the simulation, even if cells change their position. Third, the geometry of the cell is assumed to be a regular hexagon in all the simulations; nevertheless, endothelial cells adapt their shape to the local requirements. \textcolor{black}{In fact, previous works \cite{Ye2014} observed that the endothelial cell monolayer is formed by regular or non-regular hexagonal shape cells depending on the vessel curvature.} However, we have not included this effect in the simulation. 

The regulation of the endothelial monolayer plays a crucial role not only in the metastatic cascade but also in other pathologies such as pulmonary edema \cite{Collins2013} and atherosclerosis \cite{Dimmeler2004}. It is also important for the immune system as it regulates exchanges of leukocyte between the bloodstream and the surrounding tissues \cite{Strell2008}. Thus, the model we present in this work will be useful to understand these pathological and physiological conditions that crucially depend on extravasation.

\section{Acknowledgements}
Authors were supported by the European Commission (Grant no: 826494) and the Spanish Ministry of Economy and Competitiveness (Grant no: RTI2018-094494-B-C21).



  \bibliographystyle{elsarticle-num-names}




\end{document}